\documentclass[aps,pra,twocolumn,groupedaddress,showpacs,showkeys]{revtex4-1}
\usepackage[utf8]{inputenc}
\usepackage{amsmath}
\usepackage{braket}
\usepackage{graphicx}
\usepackage{pgfplots}
\usepackage{csquotes}
\usepackage{float}
\usepackage{multirow}
\usepackage{dcolumn}
\usepackage{qcircuit}
\usepackage[none]{hyphenat}
\usepackage[colorlinks=true,urlcolor=blue,citecolor=blue]{hyperref}
\usepackage{hhline}
\usepackage{subcaption}
\begin{document}
\sloppy
\title{In principle demonstration of quantum secret sharing in the IBM quantum computer}

\author{Dintomon Joy}
\email[]{dintomonjoy@cusat.ac.in}
\affiliation{Department of Physics, \\Cochin University of Science and Technology, Kochi 682022, Kerala, India
}

\author{Sabir M}
\email[]{msr@cusat.ac.in}
\affiliation{Department of Physics, \\Cochin University of Science and Technology, Kochi 682022, Kerala, India
}

\author{Bikash K. Behera}
\email[]{bkb13ms061@iiserkol.ac.in}
\affiliation{Department of Physical Sciences, \\Indian Institute of Science Education and Research Kolkata, Mohanpur 741246, West Bengal, India}

\author{Prasanta K. Panigrahi}
\email[]{pprasanta@iiserkol.ac.in}
\affiliation{Department of Physical Sciences, \\Indian Institute of Science Education and Research Kolkata, Mohanpur 741246, West Bengal, India}

\begin{abstract}
Quantum secret sharing is a way to share secret messages amongst a number of clients in a group with unconditional security.
For the first time, Hillery \textit{et al.} [Phys. Rev. A \textbf{59}, 1829 (1999)] proposed the quantum version of classical secret sharing protocol using GHZ states. Here, we implement this quantum secret sharing protocol in IBM 5-qubit quantum processor `ibmqx4' and compare the experimentally obtained results with the theoretically predicted ones. The results are analyzed through quantum state tomography technique and the fidelity of these states were calculated for different number of executions made in the device. It is concluded that the experimental results match with the theoretical values with a high fidelity.   
\end{abstract} 

\keywords{IBM Quantum Experience, Quantum Secret Sharing, Quantum State Tomography}
\maketitle

\section{Introduction \label{qss_I}}

Classical secret sharing (CSS) \cite{qss_ChenScirep2017,qss_AbulkasimIOP2016} is a method for splitting and sharing a secret message amongst n number of agents in a group, where each agent holds a share of the message. Only sufficient number of agents (let us say k) can reconstruct the secret message, while any set of k-1 or lesser number of agents can gain no information about the message. This scheme is known as the (k, n) classical threshold scheme \cite{qss_DanielPRA2000,qss_SudhirPRA2005}, which holds for all values of n and k with n $\geq$ k as independently invented by Shamir \cite{qss_ShamirACM1979} and Blakley \cite{qss_BlakleyIEEE1979}. CSS schemes are most widely used in classical cryptosystems \cite{qss_SchneierWNY1996,qss_GruskaTCPL1997} preventing any damage or stealing of secret message by an untrustworthy agent. However, by the use of advanced quantum algorithms \cite{qss_ShorASFC1994,qss_GroverACMSTC1996,qss_KarthikarXiv2017,qss_GangopadhyayQIP2018}, these schemes can be broken. In addition, CSS schemes are not perfectly secure from eavesdrop's attack, which is another major drawback \cite{qss_WangPRA2017}. 

In 1999, Hillery \textit{et al.} proposed the quantum secret sharing (QSS) \cite{qss_BuzekPRA1999}, also known as quantum information splitting (QIS) protocol using GHZ states. In this protocol, Alice shares the quantum information between two parties Bob and Charlie. Interestingly, during the process, this information gets entangled between them in such a way that, none of them can independently reconstruct the information at their location. But, later one of them can retrieve the information with the consent from the other. This protects the secret information from a possible dishonest receiver and provides unconditional security over the classical one \cite{qss_LoScience1999,qss_MayersACM2001}. A number of advantages of using QSS are mentioned as follows which will interest the readers. It can be used in the context of sharing a quantum key securely \cite{qss_WangEPJD2007} in a multi-partite scenario in the quantum money scheme \cite{qss_Wiesner1983}. QSS schemes act as efficient quantum error correcting (QEC) codes \cite{qss_ClevePRL2009,qss_MatsumotoQIP2017} that can correct erasure error \cite{qss_LuPRL2016}. It has also motivated to use the QSS scheme in a graph-theoretic protocol \cite{qss_SarvepalliPRA2012,qss_GravierTCS2015}. Recently, quantum Gauss-Jordan elimination code has been applied to QSS code to reduce the complexity of a problem \cite{qss_DiepIJTP2018}. QSS is also used in quantum dialogue to enhance the security in the protocol \cite{qss_GaoPS2018,qss_AbulkasimPS2018}. 

After Hillery \textit{et al.}, a number of theoretical \cite{qss_HsuPRA2005,qss_LiuPS2014,qss_LaiQIP2014,qss_XieIJTP2015,qss_LiPA2013,qss_TsaiSCPMA2012} schemes using entangled states \cite{qss_HelwigPRA2012,qss_LiaoQIP2014,qss_GaoIJTP2014,qss_ZhangIJTP2014,qss_ZhangPRA2005_1,qss_MassoudACPMA2012} and product states \cite{qss_GuoPLA2003,qss_YanPRA2005,qss_HanOC2008,qss_ZhangPRA2005_2,qss_ChenQIP2013,qss_WangIJTP2013,qss_WangOC2008,qss_HaoIJTP2012} have been proposed. There are various schemes available for quantum secret sharing, which are described as follows. Muralidharan and Panigrahi \cite{qss_MuralidharanPRA772008} proposed a scheme of quantum secret sharing of arbitrary single qubit and two-qubit state via teleportation which can also be used in superdense coding scheme as well. Then a scheme of QSS using multi-partite cluster states \cite{qss_MuralidharanPRA782008} was proposed. A number of schemes of QSS using multi-partite entangled states \cite{qss_BorrasJPAMT2007, qss_ChoudhuryJPAMT2009,qss_CheungPRA2009,qss_HouOC2010,qss_LiIJTP2010,qss_NieQIP2011,qss_NieSD2011,qss_MuralidharanarXiv2011} have been proposed later. Many groups have experimentally \cite{qss_Tittel,qss_Schmid,qss_Gaertner,qss_WeiOE2013} realized the QSS protocol. 
 
Even though there are numerous theoretical proposals, only very few of them got experimental realization. The recent surge in the interests among the companies like IBM, Google, Microsoft, Intel, Rigetti and D-Wave to realise quantum computing machines for commercial purpose instigated huge developments in this field. IBM has made its five-qubit quantum processors `ibmqx2', `ibmqx4' and 16-qubit processor `ibmqx5', as an open access resource \cite{qss_IBM} for the public to test and verify various theoretical protocols. Many groups were able to test various quantum computational tasks like quantum teleportation \cite{qss_Fedortchenko,qss_Mithali},  violation of Mermin inequalities \cite{qss_Alsina}, verification of entropic uncertainity relations \cite{qss_Berta}, quantum error correction \cite{qss_Devitt,qss_Takita,qss_Wooton2,qss_GhoshQIP2018}, quantum cheque \cite{qss_MoulickQIP2016,qss_BeheraQIP2017}, non-destructive discrimination of Bell states \cite{qss_Mithali2}, designing fault tolerant quantum circuits \cite{qss_Vuillot}, homomorphic encryption experiments \cite{qss_Huang}, non-Abelian braiding of surface code defects \cite{qss_Wooton}, approximate quantum adders \cite{qss_Li}, entanglement assisted invariance \cite{qss_Deffner}, simulating ising interaction \cite{qss_Hebenstreit}, comparison or quantum computing architectures \cite{qss_Linke} to name a few. 

In this paper, we discuss the implementation of quantum secret sharing scheme introduced by Hillery \textit{et al.} in the five-qubit transmon bowtie chip (`ibmqx4') and compare the obtained experimental density matrix with the theoretical one using the quantum state tomography technique. The fidelity measure is calculated to show the accuracy of the obtained results.  

In Sec. \ref{qss_III}, we briefly review the quantum secret sharing scheme of Hillery \textit{et al.} In Sec. \ref{qss_IV}, we discuss about the `ibmqx4' qubit architecture and the circuit used for implementing the HBB protocol. In Sec. \ref{qss_V}, we have discussed the quantum state tomography technique used to characterize the outputs obtained in the experiment. In Sec. \ref{qss_last} we present our conclusion and future direction of work.      

\section{Review of quantum secret sharing (QSS)  \label{qss_III}}
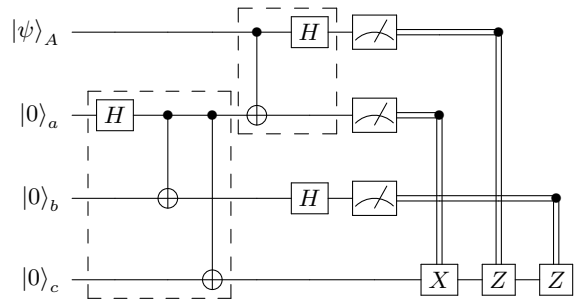
\begin{figure}
\[
\Qcircuit @C=1em @R=2em {
\lstick{\ket{\psi}_{A}}  & \qw     & \qw     &\qw      & \ctrl{1} &\gate{H} \gategroup{1}{5}{2}{6}{.7em}{--}&\meter   &\cw & \control\cw\cwx[3]    \\ 
 \lstick{\ket{0}_{a}} &\gate{H} &\ctrl{1} &\ctrl{2} & \targ    & \qw  &\meter  & \control\cw\cwx[2]  \\
 \lstick{\ket{0}_{b}} & \qw     &\targ    & \qw      & \qw   & \gate{H}  & \meter  &\cw &\cw &\control\cw\cwx[1]  \\
 \lstick{\ket{0}_{c}} & \qw     & \qw     & \targ  \gategroup{2}{2}{4}{4}{.7em}{--} & \qw      &\qw & \qw    & \gate{X} & \gate{Z} & \gate{Z}}
\]
\caption{\textbf{Quantum circuit to implement quantum secret sharing (QSS) protocol.} Here, $\ket{\psi}_A$ represents the quantum secret in Alice's possession. Qubits $a, b$ and $c$ represent the GHZ channel shared between Alice, Bob and Charlie respectively. The measurement device at the end of each qubit line measures the qubit in Z-basis. The double line after measurement represents the classical information corresponding to the output state. The first dashed box (from left to right) represents the 3-qubit GHZ state and the second one represents Bell measurement.}\label{qss_circuit}
\end{figure}

Here, we briefly review the Hillery, Buzek and Berthiaume (HBB) protocol. The Fig. \ref{qss_circuit} shows the quantum circuit representation of the HBB protocol. \\
 
\textbf{(i)} The sender Alice and the users Bob and Charlie shares a 3-qubit GHZ state $\frac{1}{\sqrt{2}}\{\ket{000}+\ket{111}\}_{abc}$ prior to the beginning of quantum secret sharing procedure \cite{qss_Karlsson}.\\
 
\textbf{(ii)} Alice wants to send an arbitrary single qubit state $\ket{\psi_{A}}=\alpha \ket{0}_A+ \beta \ket{1}_A$, in her possession to Charlie (Bob) through the method of quantum teleportation \cite{qss_Bennett}. However, Charlie (Bob) can recover the teleported state only by cooperating with Bob (Charlie)(due to quantum no-cloning theorem \cite{qss_Wootters}).\\
 
\textbf{(iii)} Alice then performs a Bell basis $\{\ket{\Psi_{\pm}}_{Aa}= \frac{1}{\sqrt{2}}(\ket{00}\pm \ket{11}) , \ket{\Phi_{\pm}}_{Aa}=\frac{1}{\sqrt{2}}(\ket{01}\pm \ket{10})\}$ measurement on the two particles $(A,a)$ in her possession and keeps the measurement result to herself. \\
 
\textbf{(iv)}  After confirming via public channel, that both Bob and Charlie have received one particle each, Alice sends her measurement result to Charlie (Bob).\\ 
 
\textbf{(v)}  Bob (Charlie) then performs a single particle measurement on his particle in X-basis $\{\ket{+},\ket{-}\}$ and sends his measurement result to Charlie (Bob).\\  
 
\textbf{(vi)} Now, Charlie (Bob) can reconstruct the teleported information by getting one bit classical information from Bob (Charlie) and the two bits earlier sent by Alice. \\
 
\begin{figure}
\[
\Qcircuit @C=1em @R=2em {
\lstick{\ket{\psi}_{A}}  & \qw     & \qw     &\qw      & \ctrl{1} &\gate{H} \gategroup{1}{5}{2}{6}{.7em}{--}&\qw   &\qw & \control\qw\qwx[3]   &\qw  &\qw   \\ 
 \lstick{\ket{0}_{a}} &\gate{H} &\ctrl{1} &\ctrl{2} & \targ    & \qw  &\qw  & \control\qw\qwx[2]  &\qw   &\qw  &\qw  \\
 \lstick{\ket{0}_{b}} & \qw     &\targ    & \qw      & \qw   & \gate{H}  & \qw  &\qw &\qw &\control\qw\qwx[1]  &\qw    \\
 \lstick{\ket{0}_{c}} & \qw     & \qw     & \targ  \gategroup{2}{2}{4}{4}{.7em}{--} & \qw      &\qw & \qw    & \gate{X} & \gate{Z} & \gate{Z} &\qw 	}
\]\caption{\textbf{The equivalent quantum circuit to implement quantum secret sharing (QSS) in IBM 5-qubit quantum processor.} The measurement device is removed and the double lines are replaced by single lines. Here, the controlled unitary operations effectively simulates the action of performing unitary operations based on the received classical information.}\label{qss_mlesscircuit}
\end{figure}
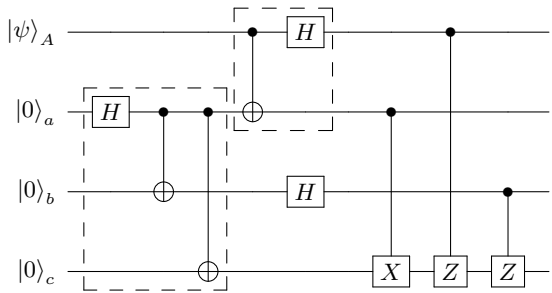

The quantum circuit given in Fig. \ref{qss_circuit} shows appropriate unitary operations to be performed by Charlie (Bob). Further this scheme allows any one of the users Bob or Charlie to recover the teleported state by interchanging their positions. In order to implement this protocol in the IBM five-qubit quantum processor, we convert this quantum circuit into an equivalent quantum circuit as given in Fig. \ref{qss_mlesscircuit} and then implement the scheme by redesigning  the circuit based on the qubit architecture of the available quantum processor, `ibmqx4'. 

\section{Experimental Realization of QSS in IBM QE \label{qss_IV}}

\begin{figure}[ht!]
   \centering
   \includegraphics[scale=0.51]{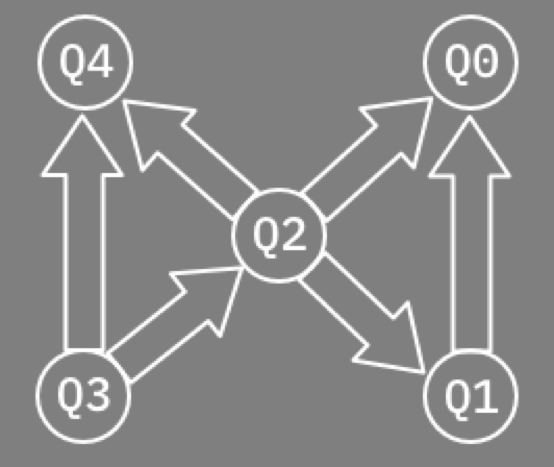}
    \caption{ibmqx4 5-qubit transmon bowtie chip (Courtesy-IBM).}
    \label{qss_chip}
\end{figure}

\begin{figure*}
\centering
    \includegraphics[width=\textwidth, height=3.5cm]{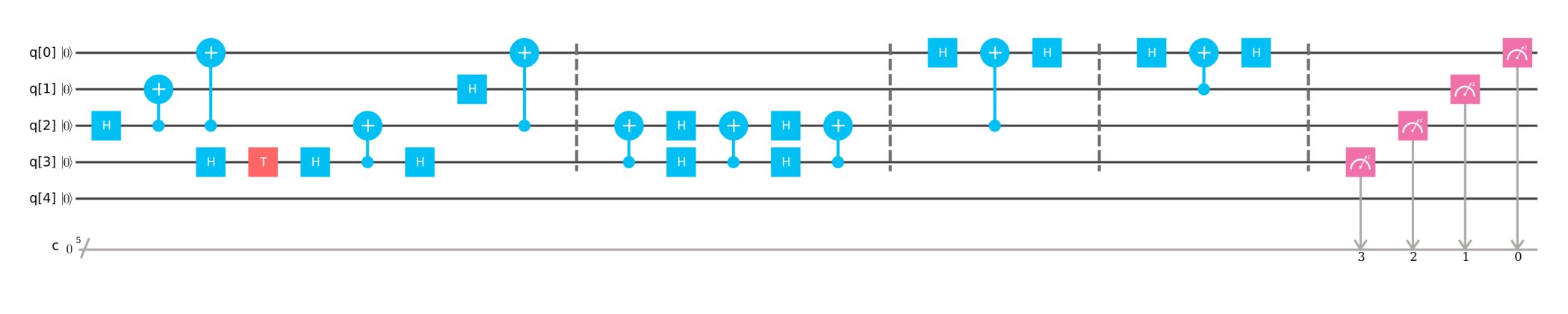}
    \caption{The circuit used for implementation of quantum secret sharing scheme in IBM quantum experience} \label{qss_ibm}
\end{figure*}

Both the five-qubit devices `ibmqx2' and `ibmqx4' are accessible through a simple online graphical user interface (GUI), allowing the users to just click, drag and drop the required quantum operations in the circuit. Alternatively, these devices and the other higher backend devices can also be easily accessed through the quantum information software Kit (QISKit), a python based software package, to implement and verify the theoretical protocols. In this work, we have implemented the HBB protocol in `ibmqx4' backend due to the better qubit architecture it provides for the problem. In order to implement HBB protocol in these processors, the measurementless circuit in Fig. \ref{qss_mlesscircuit} must be redesigned based on the qubit architecture shown in Fig. \ref{qss_chip}. Fig. \ref{qss_ibm} shows the actual quantum circuit for HBB protocol, implemented in the `ibmqx4' processor. Initially all the qubits (Q0, Q1, Q2, Q3 and Q4) are kept in the state $\ket{0}$. Then the single qubit gates are dragged in to the position in each circuit line from the toolbox according to Fig. \ref{qss_ibm}. In order to perform two-qubit operation between any two qubits, the states of each qubit must be swapped to appropriate locations. This arises due to the limitations in the qubit architecture given in Fig. \ref{qss_chip}. A useful review on accessing IBM QE and implementing quantum circuits is given by Pathak in Ref. \cite{qss_Anirban}.

The qubits Q0, Q1, Q2, Q3 and Q4 represented inside the circles are shown to be connected by an arrow in Fig. \ref{qss_chip}. The qubits near the arrow-head acts as a target qubit with respect to the control qubit at the opposite end of the arrow. The equivalent quantum circuits \cite{qss_Nielsen} are used whenever quantum gates cannot be directly implemented between any two qubits in a circuit. For example, qubit Q2 cannot be directly used as the control qubit to implement CNOT gate between Q2 \& Q3 in `ibmqx4'. Hence, it is achieved through an equivalent control reversal circuit between qubits as shown inside the SWAP operation (See Fig. \ref{qss_ibm}). The results of these circuits are obtained after executing the circuit large number of times and then collecting the probability of each outcome. The number of shots represents the number of times an experiment is executed. In this work, the results are obtained for 8192, 4096 and 1024 shots. To demonstrate the accuracy of the results, we have used the quantum state tomography technique, which is discussed in the following Sec. \ref{qss_V}.   

\subsection{Quantum State Tomography \label{qss_V}}
In HBB protocol, Alice shares the secret message $\ket{\psi_{A}}=\alpha \ket{0}_A+ \beta \ket{1}_A$ between Bob and Charlie using GHZ state $\frac{1}{\sqrt{2}}\{\ket{000}+\ket{111}\}_{abc}$ as the entangled channel. To minimize the number of gates used in the circuit, the qubits Q3, Q2, Q1 and Q0 are chosen as qubits A, a, b and c respectively. The state of Alice's secret qubit (Q3) is taken to be in a known superposition state $\ket{\psi_{A}}$, with 85$\%$ probability for finding the state in $\ket{0}$ and 15$\%$ in $\ket{1}$ state, by a successive application of $H, T$, and $H$ quantum gates,
\begin{equation}
\begin{split}
\ket{\psi_{A}}= HTH\ket{0}_{A}&=\frac{1}{2}((1+e^{i\frac{\pi}{4}})\ket{0}_{A}+(1-e^{i\frac{\pi}{4}})\ket{1}_{A})\\
\end{split}
\end{equation}
Here, the values of $\alpha$ and $\beta$ are known. The exact values for the ideal case are $\ket{0}_c$ with probability 0.8535 and $\ket{1}_c$ with 0.1464. In order to verify the HBB protocol, we must observe the same quantum state at the output qubit location Q0. However, it is known that, with a single measurement the complete information about the output state cannot be determined and also it perturbs the state during the process. Hence, we repeat the same experiment multiple times in this device to get the probability of obtaining each output state with more accuracy. In quantum state tomography, we measure the same output qubit in different basis to get the density matrix of the output quantum state. The density matrix of the system provides the complete picture of the given state and it can be obtained by using the stokes parameters \cite{qss_Altepeter}. Any single qubit state can be written as,  
\begin{equation}
\hat{\rho}=\frac{1}{2} \sum_{i=0}^{3} S_i \hat{\sigma_i}
\end{equation}
where, $\hat{\sigma_{i}}$, represents the Pauli matrices.
\begin{equation}
\sigma_0=
\begin{bmatrix}
1 & 0\\
0&1
\end{bmatrix}
;
\sigma_1=
\begin{bmatrix}
0&1\\
1&0
\end{bmatrix}
;
\sigma_2=
\begin{bmatrix}
0&-i\\
i&0
\end{bmatrix}
;
\sigma_3=
\begin{bmatrix}
1&0\\
0&-1
\end{bmatrix}
\end{equation}

$S_i$ represents the stokes parameters and its values are given by $S_i= Tr\{\hat{\sigma_i}\hat{\rho}\} $. They are related to the probability of measurement outcome ($P_{\ket{i}}$ denoting the probability of finding the given qubit in state $\ket{i}$) by $S_0=P_{\ket{0}}+P_{\ket{1}}$, $S_1=P_{\ket{+}_X}-P_{\ket{-}_{X}}$, $S_2=P_{\ket{+}_Y}-P_{\ket{-}_Y}$ and $S_3=P_{\ket{0}}-P_{\ket{1}}$. The parameter $S_0$ always equals unity in order to conserve probability. The states $\{\ket{+}_{X}, \ket{-}_{X}\}$, $\{\ket{+}_{Y}, \ket{-}_{Y}\}$ and $\{\ket{0}, \ket{1}\}$ represent the eigen states of Pauli X, Y and Z matrices respectively. By calculating these unique stokes parameters, we can specify the position of an arbitrary single qubit state in the Bloch sphere.  

We have executed the circuit given in Fig.~\ref{qss_ibm} in `ibmqx4' and made measurements on the qubit Q0 in different bases for different number of shots, to find the probability distribution of the results. By default the measurement in IBM 5-qubit quantum processor is made in Z-basis. To make a measurement in X-basis, a $H$ gate is inserted just before the measurement operator. Similarly, to measure the qubit in Y-basis, $S^{\dagger} H$ gates are placed before measurement operator. Then the probabilities of getting particular measurement results are found, to construct the stokes parameters. In effect, we measured the state of the qubit Q0 in different bases and used stokes parameters to reconstruct the density matrix of the system.
 
\begin{figure}
\centering
   \includegraphics[scale=0.45]{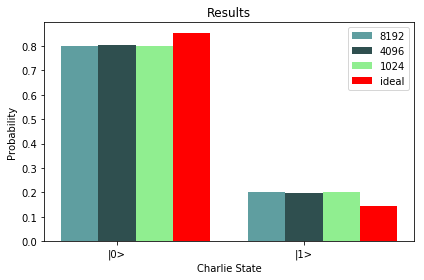}
    \caption{The bar plot comparison of the results obtained from real device are colored blue, grey and green for shots 8192, 4192 and 1024 respectively. For the ideal case it is colored red.}
    \label{qss_results}
\end{figure}

Tables \ref{qss_ibmqx4} and \ref{qss_simu} shows the output state obtained through the real device `ibmqx4' and the simulator respectively. Fig. \ref{qss_results} shows pictorial comparison of the results obtained in `ibmqx4' vs the ideal case. 

\begin{table}
\begin{center}
\caption{Run results}
\begin{tabular}{c c c} 
 \hline
 \hline
 No. of Shots & Probability of $\ket{0}_C$ & Probability of $\ket{1}_C$ \\ 
 \hline
  8192& 0.800& 0.200 \\ 
  4096& 0.803& 0.197 \\ 
  1024& 0.798& 0.203 \\ 
 \hline
 \hline
\end{tabular}
\label{qss_ibmqx4}
\end{center}
\end{table}

\begin{table}
\begin{center}
\caption{Simulated results}
\begin{tabular}{c c c} 
 \hline
 \hline
 No. of Shots & Probability of $\ket{0}_C$ & Probability of $\ket{1}_C$ \\ 
 \hline
  8192& 0.853& 0.147 \\ 
  4096& 0.860& 0.139 \\ 
  1024& 0.850& 0.151 \\ 
 \hline
 \hline
\end{tabular}
\label{qss_simu}
\end{center}
\end{table}

\begin{table}
\begin{center}
\begin{tabular}{c c c c c c} 
 \hline
 \hline
 No. of Shots & $S_0$ & $S_1$ & $S_2$ & $S_3$ & Fidelity \\ 
 \hline
  8192& 1.0000& 0.1020 & 0.0210& 0.6000 & 0.8284 \\ 
  4096& 1.0000& 0.0920 & 0.0810& 0.6060 & 0.8291\\ 
  1024& 1.0000& 0.1650 & 0.0310& 0.5950 & 0.8267 \\ 
 \hline
 \hline
\end{tabular}\caption{Stokes parameters and fidelity obtained for different number of shots.}
\label{qss_fidelity}
\end{center}
\end{table}

We calculate a measure called fidelity to estimate the closeness between the theoretical and experimental density matrices. Fidelity \cite{qss_Nielsen} is calculated using the formula given by Eq. \eqref{qss_Eq4}.  
\begin{equation}
F(\rho^{T}, \rho^{E})= Tr\{\sqrt{\sqrt{\rho^{T}}\rho^{E}\sqrt{\rho^{T}}}\}
\label{qss_Eq4}
\end{equation}

The theoretical density matrix ($\rho^T$) and experimental density matrix ($\rho^E$) obtained from the experimental results (See Table \ref{qss_fidelity}) are given by Eqs. \eqref{qss_Eq5} \& \eqref{qss_Eq6}. 
\begin{equation}
\rho^T = \left(\begin{array}{cc} 0.8535 & 0.0000 \\ 0.0000 & 0.1464 \end{array}\right)+ i \left(\begin{array}{cc} 0.0000 & -0.3535  \\ 0.3535 & 0.0000  \end{array}\right)
\label{qss_Eq5}
\end{equation}
\begin{equation}
\rho^E= \left(\begin{array}{cc} 0.8000 & 0.0510 \\ 0.0510 & 0.2000 \end{array}\right)+ i \left(\begin{array}{cc}  0.0000 & -0.0105 \\ 0.0105 & 0.0000 \end{array}\right)
\label{qss_Eq6}
\end{equation}

We found that the fidelity of output state in our experiment for 8192 shots turns out to be 0.82.

\section{Conclusion}
\label{qss_last}
We have successfully implemented HBB protocol in IBM 5-qubit quantum computer, verified the results through quantum state tomography and the accuracy through fidelity parameter. We found that the `ibmqx4' performs the implementation of the HBB protocol by providing a fidelity of 0.82. In near future, we hope to extend the same protocol to implement a binary voting protocol in `ibmqx5' backend. 

\section*{Acknowledgments}
B.K.B. acknowledges the support of Inspire Fellowship awarded by DST, Govt. of India. The authors are extremely grateful to IBM team and IBM QE project. The discussions and opinions developed in this paper are only those of the authors and do not reflect the opinions of IBM or IBM Quantum Experience team.

\end{document}